# Single-filament Composite MgB$_2$/SUS Ribbons by Powder-In-Tube Process

K. J. Song, N. J. Lee, H. M. Jang, H. S. Ha, D. W. Ha,

S. S. Oh, M. H. Sohn, Y. K. Kwon, and K. S. Ryu

Korea Electrotechnology Research Institute, Changwon, Kyungnam, 641-600, Korea

We report the successful fabrication of single-filament composite MgB$_2$/SUS ribbons, as an ultra-robust conductor type, employing the powder-in-tube (PIT) process, by swaging and cold rolling only. The remarkable transport critical current ($I_c$) of the non-sintered MgB$_2$/SUS ribbon has observed, as an unexpected result. Transport critical currents $I_c \sim 316$ A at T = 4.2 K and $I_c \sim 82$ A at T = 20 K were observed at self-field, for the non-sintered composite MgB$_2$/SUS ribbon. In addition, the persistent current density $J_p$ values, that were estimated by Bean formula, were more than $\sim 7 \times 10^5$ A/cm$^2$ at T = 5 K, and $\sim 1.2 \times 10^5$ A/cm$^2$ at T = 30 K, for the sintered composite MgB$_2$/SUS ribbon, at H = 0 G.

The magnesium diboride ($MgB_2$) that recently was discovered at $T_c$ = 39 K superconductivity[1] has seemed to be the sweetest gift in the superconductivity world. This simple binary intermetallic $MgB_2$ has led to remarkable interest in both a fundamental understanding of its superconductivity[2-4] and technological applications.[5-8] The bulk $MgB_2$ superconductor has showed both the strong-link current flow between grains[9] and the stability in time of supercurrents,[10] in contrast with both weak-link problem and giant flux creep (a rapid decay of critical currents in the mixed state) of high-$T_c$ cuprate superconductors (HTS). In addition, we are encouraged by the rapid and reliable compound synthesis as well as the commercial availability $MgB_2$ powders for the technological applications such as $MgB_2$ wires/ribbons. The Bi-cuprates were the first to be fabricated into wires/ribbons in lengths with enough texture to provide good intergrain current transport. Their application, however, is restricted because they have a low lying irreversibility line $H_{irr}(T)$. In developing $MgB_2$ wires/ribbons, recently Jin et al.[8] showed the fabrication of dense and metal-clad $MgB_2$ superconductor wire with transport $J_c$ of greater than 36,000 A/cm$^2$. The critical current density $J_c$ evaluated by magnetization measurement was reported to be $10^5 \sim 10^6$ A/cm$^2$ in dense bulk $MgB_2$,[2,10] $\sim 5 \times 10^5$ in dense $MgB_2$ wire,[5] and $\sim 6 \times 10^6$ in thin films of $MgB_2$,[6] at liquid helium temperature. To progress in the development of $MgB_2$ wires/ribbons with high $J_c$ for industrial applications, it will be necessary to find optimum conditions for fabrication. Because of hampering properties such as the volatility and reactivity of magnesium, however, forming practical products, $MgB_2$ wires/ribbons, presents some challenges for us.

In this letter, we report the successful fabrication of single-filament composite $MgB_2$/SUS ribbons, as an ultra-robust conductor type, employing the powder-in-tube (PIT) process. In addition to describing the fabrication by swaging and cold rolling only, we report the superconducting results of both a sintered and a non-sintered single-filament composite $MgB_2$/SUS ribbon. The SUS-304 used as a metal cladding is an austenitic stainless steel. It is very critical point to pick a good complementary clad material for the $MgB_2$ superconductor. The synthesized or commercially available $MgB_2$ powders are fine grain polycrystalline superconductors that are mechanically hard, a bit brittle, and crumbling. When we fabricate the metal-clad $MgB_2$ wires/ribbons, hard, but ductile and malleable metals are essential. They have to play a key role as a diffusion barrier for the volatile and reactive magnesium. Furthermore, they should be strong clad materials to form dense and tiny grain superconducting wires/ribbons, due to pressure on processes such as swaging and rolling.

After careful consideration of many possible materials, we decided to use SUS-304 (a type of stainless steel) tubes as a clad metal for $MgB_2$ ribbons as well as a diffusion barrier for sintering at high temperature near 900º C. The SUS-304 tube had an outside diameter of 10.5 mm, a wall thickness of 2.4 mm, and a length of 150 mm. First, one end of the SUS-304 tube was sealed by plugging it with a short tight-fitting SUS-304 rod. The commercially available $MgB_2$ powder (from Alfa Aesar) was then filled and packed inside the tube by pressing the SUS-304 rod with a hammer, in argon atmosphere. The other end of the tube was sealed by plugging it with a rod, as well. The $MgB_2$ powder inside the tube had an initial packing density of approximately 1.47 g/cm$^3$ (sintering sample) and 1.64 g/cm$^3$ (non-sintering sample). The tube filled with $MgB_2$ powder was then swaged down to a 2 ~ 3 mm diameter rod step by step, with cooling the composite tube by water. This swaging was followed cold rolling of many steps, each with a thickness reduction of ~ 10 %. Finally, the single-filament composite MgB$_2$/SUS ribbons reached a ribbon geometry with a thickness of ~ 0.8 mm, a width of ~ 5 mm, and a length of ~ 1 m. Without cutting, one of the prototype ribbons (initial packing density of 1.47 g/cm$^3$) was given a sintering treatment at 900$^\circ$ C for 1 hour in an argon atmosphere, with additional sealing both ends of ribbons by Ar-arc soldering. We estimate that the loss of mass during sintering is almost zero, except very small areas of reaction between MgB$_2$ and SUS-304 tube.

The scanning electron microscopy (SEM) and electron probe X-ray micro-analyzer (EPMA-1600) investigations for both the sintered and the non-sintered MgB$_2$/SUS ribbons are shown in Fig. 1-(a) ~ (d). Fig. 1-(a) and -(b) show the typical transverse cross sections. In Fig. 1-(c) and -(d) are shown the EPMA line profiles of the regions marked in Fig. 1-(a) and -(b), for both the sintered and the non-sintered MgB$_2$/SUS ribbons, respectively. The superconductor core cross section areas of the composite MgB$_2$/SUS ribbons that show irregular forms are about 0.67 and 1.13 mm$^2$ for the sintered and the non-sintered one, with the superconductor filling factor of about 19.5 and 30.0 %, respectively. Usual SUS-304 consists of 20 % Cr, 10 % Ni, 2 % Mn, 1 % Si, very small potions of other elements, and the remainder is Fe. No evidence of significant diffusion and reaction between MgB$_2$ and SUS-304 tube is found in the EPMA spectra. As shown in Fig. 1-(c) and (d), where the numbers are the counts of each element for the line profiles, however, there are a bit reacted traces with oxygen into MgB$_2$ core for both the composite ribbons. An impurity phase, such as a MgO, might be generated inside the superconductor MgB$_2$ core during fabrication.

For the resistivity ($\rho$) and the transport critical current ($I_c$) measurements, we cut segments of both sintered and non-sintered composite MgB$_2$/SUS ribbons with the length of 4 ~ 5 cm long. Their characteristic curves, R(T) as well as V(I), were measured by using a standard dc four-probe method. The normalized resistivity curves shown in Fig. 2 for both prototype composite ribbons were obtained with directly soldered voltage contacts separated by the distance of ~ 10 mm, using a 30 mA current. The normalized resistivity curves at self field, measured over a wide temperature range from 300 to 10 K, show a sharp superconducting transition at $T_c$ (onset) = 38.5 K and 36 K for the sintered and the non-sintered one, respectively. The sintered one was found to slightly increase the $T_c$ as compared to the non-sintered one that exhibited about 2.5 K lower $T_c$. The non-sintered one, however, presents much sharper transition and higher $T_c$ than those of the pellet type sample, that was pressed with 3 GPa without any heat treatment, of commercially available $MgB_2$ powder.[11] The transport critical current of the non-sintered composite $MgB_2$/SUS ribbon was obtained from V(I) curve, employing AR4800 (32-channel, Yokogawa), with directly soldered voltage contacts separated by the distance of ~ 10 mm, by sweeping the current of up to ~ 400 amperes. On determining $I_c$ with the ~ 1 $\mu$V/cm criterion, we have taken care of the superimposed voltages of the superconductor $MgB_2$ core and the clad SUS-304. As shown in Fig. 3, transport critical currents $I_c$ ~ 316 A at T = 4.2 K and ~ 82 A at 20 K were measured at self-field for the non-sintered composite $MgB_2$/SUS ribbon. By using a superconductor $MgB_2$ core cross section areas of ~ 1.13 $mm^2$, we can estimated the transport critical current densities of ~ $2.8 \times 10^4$ and of ~ $7.3 \times 10^3$ $A/cm^2$ at 4.2 K and 20 K, respectively. In addition, we measured the field dependence of $I_c$ for the non-sintered one at T = 4.2 K, as shown in the inset of Fig. 3. All of data followed an exponential fall of $I_c$ with H, except H = 3 T. The deviation at H = 3 T seems to be due to shielding effect of the clad SUS. It, however, is hard to compare directly the transport critical current of both the sintered and the non-sintered composite $MgB_2$/SUS ribbons. Not only, do they have different superconductor filling factors, but also different reacted traces on the boundaries, shown in Fig. 1-(a) and -(b), between the core $MgB_2$ and the clad SUS of the composites.

For magnetization study, we cut the ribbon into pieces with dimensions of about $0.5 \times 0.5 \times 0.08$ mm$^3$. Unfortunately, it was very hard to measure the superconductive magnetization of the non-sintered composite $MgB_2$/SUS ribbon, because of the ferromagnetism induced in the cladding by swaging and rolling. The ferromagnetic behavior, however, faded and the SUS clad became less magnetic while sintering the composite $MgB_2$/SUS ribbons at 900° C for 1 hour. We, therefore, can measure and study the magnetization for the composite structures of the sintered composite $MgB_2$/SUS ribbon. Using a PPMS-9 (Quantum Design), the isothermal magnetization M(H) was measured at temperatures T between 5 and 50 K in magnetic fields up to 6 T. Values of magnetization M were corrected for the background M, measured above $T_c$. The inset in Fig. 4 shows the zero-field-cooled curve for H perpendicular to platelet plane. Measurement of the low-field susceptibility $\chi$ in H = 10 G revealed $T_c$'s of 38.5 K for H perpendicular to the platelet plane. The persistent current density $J_p$ was obtained from the irreversible magnetization using the Bean critical state model. The Bean formula to estimate the persistent current density is $J_p \sim 30 \Delta M/d$ with J in A/cm$^2$, where $\Delta M$ is the difference of magnetization M between the increasing field and the decreasing field branches, in units of emu/cm$^3$ = G, and d is the transverse width of the superconductor core, in the sintered ribbon. Fig. 4 shows the persistent current density $J_p$ versus magnetic field H for each temperature. As shown in Fig. 4, $J_p$ values were estimated to be more than $\sim 7 \times 10^5$ A/cm$^2$ at H = 0 T, and $\sim 6 \times 10^5$ A/cm$^2$ at 1 T, for the sintered ribbon at T = 5 K. In addition, $J_p \sim 1.2 \times 10^5$ at H = 0 and $\sim 2 \times 10^4$ at 1 T were estimated at 30 K.

Let us recall the remarkable transport critical current ($I_c$) for the non-sintered composite $MgB_2$/SUS ribbon, as an unexpected result. During the swaging and cold rolling processes, the superconductor core has suffered a deformation by high pressure on every step, because of the hard clad SUS-304. The composite $MgB_2$/SUS ribbons, therefore, can maintain the fine $MgB_2$ particles in intimate contact inside hard SUS-304 tube on every swaging and rolling cycle than the starting $MgB_2$ powder. In contrast with the prominent weak-linkage between grains in high-$T_c$ cuprate superconductors, polycrystalline forms of superconducting $MgB_2$ have shown strongly linked current flow between misoriented grains, as observed by Larbalestier, et al.[9] Therefore, the finer superconductor $MgB_2$, that has the striking feature of strong-linkage between grains, might be a factor for vortex pinning enhancement. On the contrary, there is a possibility of some reaction between the $MgB_2$ and the SUS-304 clad at high temperature. So, some non-superconducting phases, such as MgO or others, could be a barricade to current flow between grains. The composite $MgB_2$/SUS ribbons, however, offer a positive vision of reaching much higher persistent critical densities, by better controlling of contamination and stabilizing of the superconductor during fabrication, of the composite $MgB_2$/SUS ribbons or other metal or non-metal clad/sheath $MgB_2$ ribbons. There are still large open areas to understand and to improve the superconductivity for metal-clad $MgB_2$ wires/ribbons.

In summary, we successfully have fabricated a single-filament composite $MgB_2$/SUS ribbon, as an ultra-robust conductor type, employed the powder-in-tube (PIT) process that was swaging and cold rolling only. In addition, we report a remarkable transport critical current ($I_c$) of the non-sintered composite $MgB_2$/SUS ribbon, as a striking result. The transport critical current $I_c \sim 316$ A at T = 4.2 K and $I_c \sim 82$ A at T = 20 K were observed at self-field, for the non-sintered one. Sharp superconducting transitions at $T_c$ (onset) = 38.5 K and 36 K were observed from resistivity curves for the sintered and the non-sintered composites, respectively. The sintered composite was found to have a slightly higher $T_c$ as compared to the non-sintered composite one that exhibited a $T_c$ about 2.5 K lower. On the other hand, for the sintered composite one, the persistent current density $J_p$ values, that were estimated by Bean formula, were more than $\sim 7 \times 10^5$ A/cm$^2$ at H = 0 G, and $\sim 6 \times 10^5$ A/cm$^2$ at H = 1 T, at T = 5 K, respectively. At 30 K, $J_p \sim 1.2 \times 10^5$ at H = 0 G and $J_p \sim 2 \times 10^4$ at 1 T, as well. We are still optimizing the processes for forming single or multi-filament composite $MgB_2$/SUS ribbons.

The author (K. J. Song) wishes to acknowledge valuable discussions with J. R. Thompson. This work is supported by the Korea Electrotechnology Research Institute.

Correspondence and requests for materials should be addressed to K. J. Song (e-mail: kjsong@keri.re.kr)

References:

1. Jun Nagamatsu, *et al. Nature* **401**, 63-64 (2001)

2. D. K. Finnemore, *et al. Phys. Rev. Lett.* **86**, 2420-2422 (2001)

3. S. L. Bud' ko, *et al. Phys. Rev. Lett.* **86**, 1877-1880 (2001)

4. W. N. Kang, *et al. Preprint at* http://xxx.lanl.gov *Cond-mat/0102313* (2001)

5. P. C. Canfield, *et al. Phys. Rev. Lett.* **86**, 2423-2426 (2001)

6. W. N. Kang, *et al. Science*, 12 April 2001 (10.1126/Science 1060822) or *Preprint at* http://xxx.lanl.gov *Cond-mat/0103179* (2001)

7. Y. Bugoslavsky, *et al. Nature* **410**, 563-565 (2001)

8. S. Jin, *et al. submitted to Nature*, *Preprint at* http://xxx.lanl.gov *Cond-mat/0104236* (2001)

9. D. C. Labalestier, *et al. Nature* **410**, 186-189 (2001)

10. J. R. Thompson, *et al. Supercond. Sci. Technol.* **14**, L19 (2001)

11. C. U. Jung, *et al. Preprint at* http://xxx.lanl.gov *Cond-mat/0102383* (2001)

Figure Captions:

Fig. 1. SEM and EPMA images of both the sintered and the non-sintered single-filament composite $MgB_2$/SUS ribbons; typical transverse cross sections of both (a) the sintered one and (b) the non-sintered one, and EPMA line profiles of both (c) the sintered and (d) the non-sintered ribbons. The numbers for the EPMA spectra shows the counts for each element.

Fig. 2. Temperature dependence of the normalized resistivity of single-filament composite $MgB_2$/SUS ribbons. The lower inset shows a magnified view near $T_c$. The superconducting transitions are observed at 38.5 K and 36 K for the sintered (circle) and non-sintered (square) composite ribbons, respectively.

Fig. 3. Voltage versus Current characteristics for a non-sintered single-filament composite $MgB_2$/SUS ribbon. Defining $I_c$ with a ~1 µV/cm criterion, the transport critical current is $I_c$ ~ 316 A at 4.2 K (open circle) and $I_c$ ~ 82 A at 20 K (crossed circle). The upper inset shows the field dependence of $I_c$ for a non-sintered composite at T = 4.2 K. The results are well described by an exponential decrease of $I_c$ with H, except H = 3 T.

Fig. 4. The persistent current density $J_p$ (from Bean critical state formula) versus magnetic field (H), with magnetic field perpendicular to platelet plane, for a sintered single-filament composite $MgB_2$/SUS ribbons at T = 5 K to 35 K. The lower inset shows the ZFC curve for H (= 10 G) perpendicular to platelet plane, too. The measurement reveals $T_c$'s of 38.5 K for a sintered composite ribbon.

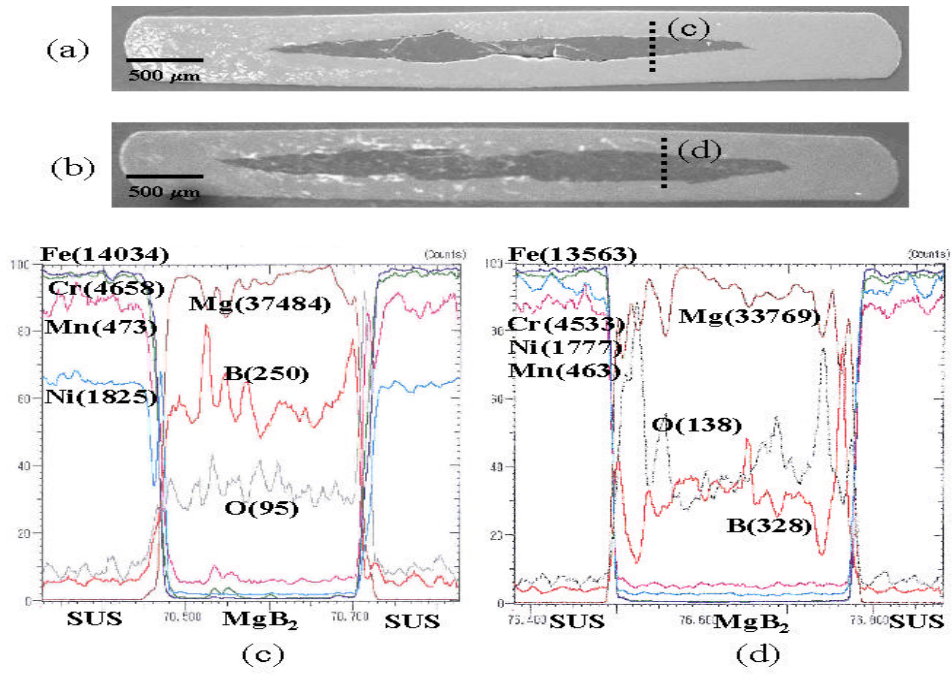

Fig. 1, K. J. Song et al.

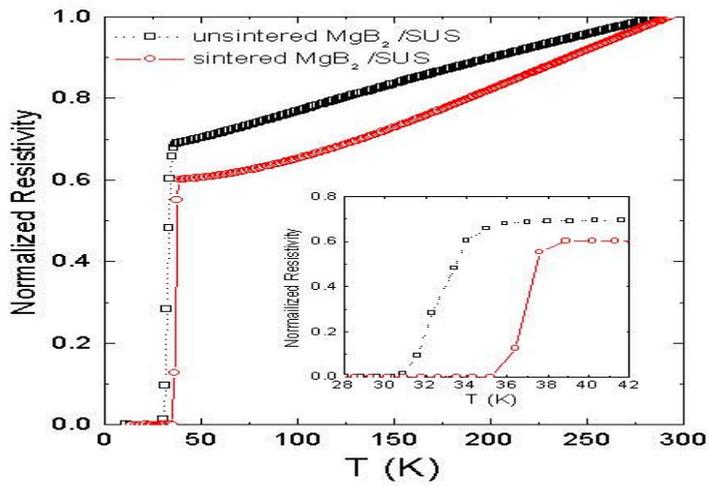

Fig. 2, K. J. Song et al.

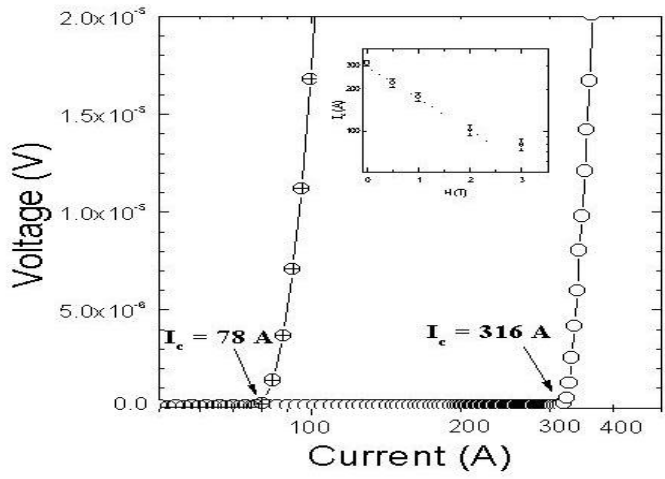

Fig. 3, K. J Song et al.

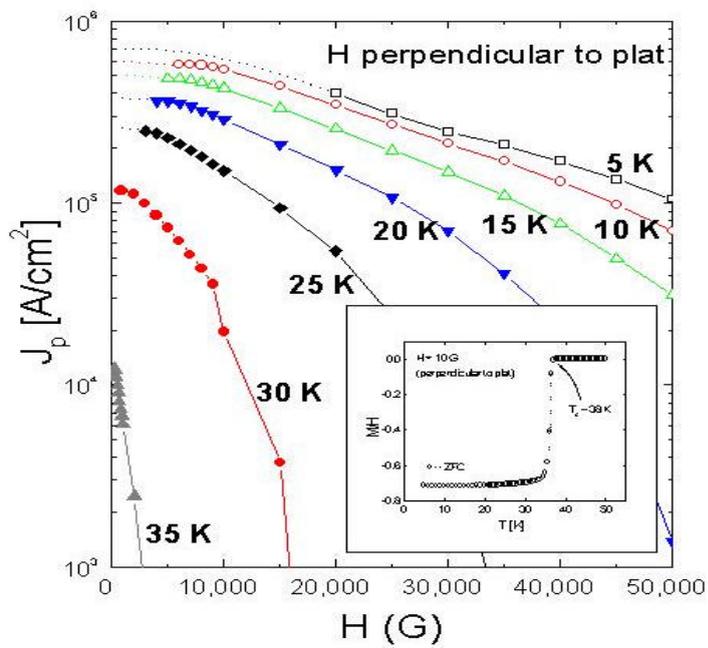

Fig. 4, K. J. Song et al.

12